\documentstyle[12pt,aps,epsf]{revtex}

\input{epsf}
\begin{document}

\title{Backscattering of light by a finite two-dimensional plate
in the region of excitonic absorption.}
\author{Kozlov G.G.}

\maketitle
 \hskip20pt All-Russia research
 center "Vavilov State Optical Institute",
 St. Petersburg, Russia
\vskip20pt
\hskip100pt {\it e}-mail:  gkozlov@photonics.phys.spbu.ru
\vskip20pt
\begin{abstract}
It is shown that  the light scattered by a finite 2D crystal strip
in the region of an exciton resonance should display strong backscattering.
  This phenomenon is related to the exciton reflection
 from the edge of the strip.
\end{abstract}
\section{The essence of the effect}

  Recent achievements  in fabrication of quasi-two-dimensional
  crystal structures - quantum wells, Langmuir-Blodgett  films -
  give rise to  growing interest to the spectroscopy of this
  objects. Usually optical excitations of these systems can be
  treated as excitons. This results in non-local  character of optical
  susceptibility and strong effects of spatial dispersion.
  One of
 the effects of this kind  is  described below. Let
  us consider a finite  2D crystal strip (Fig.1) in XY
  plane which is infinite along Y-direction and occupies the
  interval x$\in[-l,l]$ along X-direction. A monochromatic
  wave (incident wave) falling on this structure from the upper half-space at
  some angle produces an exciton with the wave vector defined by the
  in-plane wave vector component of the incident wave.  In the
  case under consideration (Fig.1) this exciton will propagate
  in the left direction.
The polarization related to this exciton produces the scattered
field giving rise to conventional reflected and transmitted waves.
{\it But due to the finiteness of the strip in the X-direction,
the exciton reflected from the left edge of the strip appears.
This exciton propagates in the right-hand direction, and the
polarization related to this exciton produces a scattered wave
counter-propagating with respect to the incident wave in the upper half-space and
a symmetrical refracted wave in the lower half-space.}
So, we see that the light scattering on a finite
crystal strip in the spectral region of exciton
resonance is accompanied by a non-trivial backscattering
in the upper half-space and by a symmetrical refraction in the lower half-space.
This is the essence of the effect suggested.
In the extended version of this letter we study a simple
 model of  2D exciton which allows us to calculate the excitonic Green's
 function and non-local susceptibility of a 2D finite crystal strip.
 Given the non-local susceptibility, we can calculate the scattered
 field for the case of
   the finite strip under consideration. The result of this calculation
  is presented on Fig.2.

\begin{figure}
\epsfxsize=400pt
\epsffile{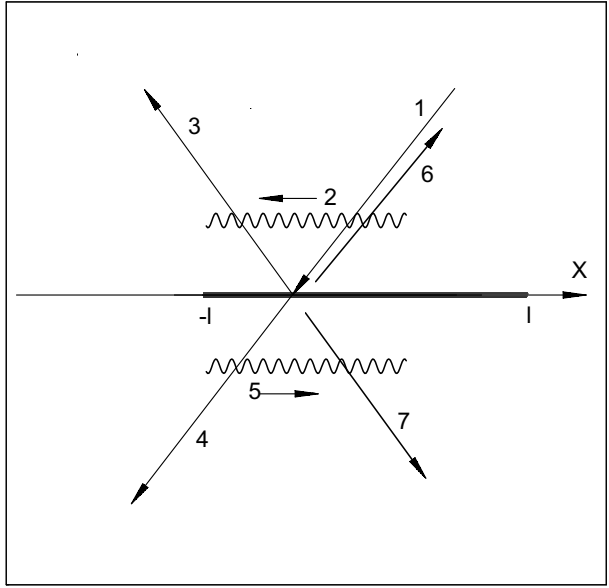}
\caption{Scattering of the plane wave by the
two-dimensional crystal strip with exciton.
1. -- incident plane wave,
 2 -- exciton produced by the incident wave
 running from the left to the right and the
 corresponding polarization.
3,4 -- the conventional reflected and transmitted
waves, 5 -- the wave of polarization related to the
exciton reflected from the left edge of the strip.
6,7 -- scattered waves produced by the reflected exciton.}
\end{figure}
\begin{figure}
\epsfxsize=400pt
\epsffile{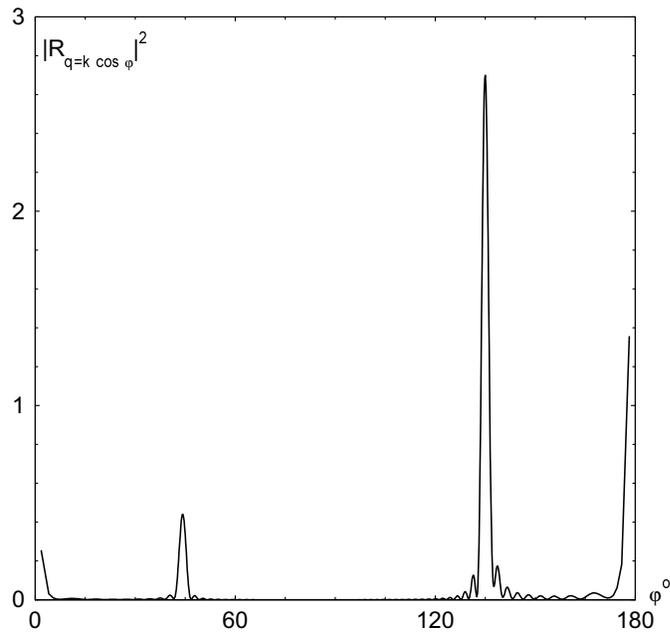}
\caption{Angular dependence of the scattering
for the case of 2D crystal strip in the spectral region of the exciton resonance.
The calculation is performed for the incident wave falling at $45^0$
 with respect to the plane
of 2D crystal. It is seen that the conventional reflection (maximum at $135^0$)
 is accompanied by the backscattering peaked at $45^0$.}
\end{figure}
\end{document}